\documentstyle[twoside,fleqn,espcrc2,epsf]{article}

\newcommand{\AmS}{{\protect\the\textfont2
  A\kern-.1667em\lower.5ex\hbox{M}\kern-.125emS}}

\newcommand{\be}{\begin{equation}}
\newcommand{\ee}{\end{equation}}
\newcommand{\ba}{\begin{eqnarray}}
\newcommand{\ea}{\end{eqnarray}}

\title{B spectroscopy using all-to-all propagators}

\author{J. Peisa
and 
C. Michael\address{Department of Mathematical Sciences, University of Liverpool, Liverpool L69 3BX, UK}, {\em UKQCD Collaboration}
}

\begin{document}

\begin{abstract}
We measure the ground and excited states for B mesons in the static limit
using maximally variance reduced estimators for light quark
propagators. Because of the large number of propagators we are able to
measure accurately also orbitally excited $P, D$ and $F$ states. We
also present some results for $\Lambda_b$.
\end{abstract}

\maketitle

\section{INTRODUCTION}

Usually one calculates the light quark propagators needed for hadronic
observables in lattice QCD by exact inversion from one source. It is
necessary to iterate almost to machine precision to avoid introducing
bias. This is clearly wasteful, as the variation from one gauge to
another can be large. The situation is even more severe for static quarks,
when one gets only one number per inversion. 

Stochastic estimators for an inverse of a positive definite matrix $M$
are easily obtained from 
\be
M_{ij}^{-1} = \left< \phi_i^* \phi_j \right> = {1\over Z}\int d\phi 
e^{-{1\over 2}\phi^*M\phi}.
\ee
Unfortunately the
Wilson-Dirac fermion matrix $Q$ is positive definite only for
extremely small values
of hopping parameter $K$. This problem can be solved by taking
$M=Q^\dagger Q$, and calculating the propagator from
\be
G_{ij} = Q^{-1}_{ji} = \left<(Q_{ik}\phi_k)^*\phi_j\right>.
\label{propag}
\ee

This stochastic estimate can be used to calculate any hadronic
observable, but because one is usually interested in exponential decay of
correlators at large $T$ values, the variance coming from stochastic
inversion will kill the signal rather quickly. Therefore one would
prefer to use improved operators; this can be achieved using variance
reduction.

\section{MAXIMAL VARIANCE REDUCTION}

One suggestion is to use local multi-hit variance reduction for $\phi$
fields \cite{divitiis}; this is very easy to implement and provides
significant error reduction. However, as the action is quadratic in
$\phi$ one should be able to implement a method that takes into account
not only the nearest neighbours of $\phi$, but all fields inside some
given region $P$.

If the scalar fields inside region $P$ and on the boundary of $P$ are
called $\phi$ and $s$ respectively, one can obtain maximally variance
reduced estimate for $\phi(x)$ by 
\ba
v_i&=&{1\over Z} \int D\phi\,\, \phi_i \times \nonumber\\
&& \exp(-{1\over 2} \phi_j^*\hat{M}_{jk}\phi_k + \phi_j^*\bar{M}_{jk}s_k +h.c.)
\label{maxred}
\ea
where we have also distinguished the elements of M connecting
$\phi_i$ to only fields inside $P$ ($\hat{M}$) and those connecting
$\phi_i$ to the boundary ($\bar{M}$). The integral (\ref{maxred})
is Gaussian and one obtains easily
\be
v_i = -\hat{M}_{ij}^{-1} \bar{M}_{jk} s_k.
\ee
We will call $v_i$ the maximally variance reduced stochastic estimator
for $\phi_i$.

To form propagators, we need a product of two $\phi$ fields. This
means that it is not sufficient to improve only one field within $P$,
but one must choose two disjoint regions $P$ and $R$ and solve for two
variance reduced fields $v$ and $w$ respectively. Then one can apply
(\ref{propag}) to obtain
\be
G_{ij} = \left< (Q_{ik}v_k)^*w_j \right>
\label{impprop}
\ee
with no bias.

\section{B MESONS AND $\Lambda_b$}

Consider mesons containing one infinitely heavy quark $Q$ and one light
quark $q$. Such a system is obtained in leading order heavy quark
effective theory HQET \cite{hqet} for the B meson. This system is particularly
suitable for stochastic inversion as usually one is only able to
obtain one number per inversion.

The expectation value one has to calculate contains simply one light
propagator and the product of gauge links $U$. This is easily obtained
from equation (\ref{impprop}). In addition one can use smearing (or fuzzing) to
create different sources; in particular it is possible to construct
operators corresponding to orbitally excited states by using techniques
from \cite{lm}.

It is also possible to construct observables with more than one light
quark. $\Lambda_b$ in leading order HQET contains one infinitely heavy
quark and two light quarks. This is also easily obtained from
stochastically improved fields $v, w$ but a little care is needed. One
way to grasp the subtlety is to imagine that there are two quarks with
different flavours. Then one has to split the sum over samples $\alpha = 1
\ldots N$ into subsets for each flavour. If these subset are
independent, one obtains propagators of each flavour with no bias. In
practice if the stochastic estimators $\phi^\alpha$ are independent of
$\alpha$, one can calculate the required propagators from
\be
G_{ij}^\alpha G_{i'j'}^\beta = \sum_{\alpha \ne \beta} 
(Q_{ik}v_k^\alpha)^* (Q_{i'k'}v_{k'}^\beta)^* w_j^\alpha w_{j'}^\beta.
\label{lambda}
\ee
The $\Lambda_b$ correlator is then easily constructed by multiplying two light
quark propagators from equation (\ref{lambda}) by the gauge
links corresponding to a heavy quark propagator.

\section{RESULTS}

\begin{figure}
  \centering
  \leavevmode
  \vspace{-2cm}
  \epsfxsize = 8cm
  \epsfbox{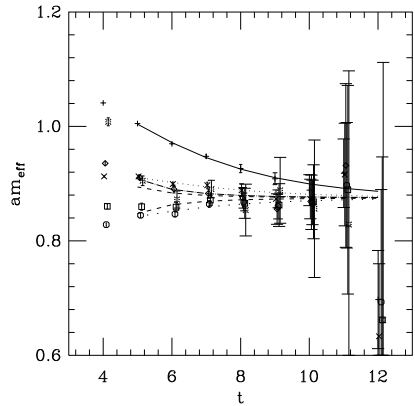}
  \vspace{-1cm}
  \caption{\protect Effective mass plot for $S$-wave with hopping
parameter $K_1=0.14077$ corresponding to roughly strange quark mass.}
  \label{figure:effmass}
\end{figure}

We have performed simulations with tadpole improved quenched clover action
($c_{SW} = 1.57$), with $\beta=5.7$. The lattice size we use is rather small,
$12^3\times 24$. We work with two different hopping parameters $K_1 =
0.14077$ and $K_2 = 0.13843$. The same values of hopping parameters
are used also in \cite{scale}, and we use those results also to set
our scale. The values $K_1$ and $K_2$ correspond to the strange quark
and twice the
strange light quark masses respectively.

We use 20 gauge configurations, and in each of them we calculate 24
stochastic samples for each hopping parameter. The stochastic update
consisted of 5 overrelaxation steps followed by one heat bath. Measurements
were taken after every 25 combined OR/HB sweeps and the system was
thermalised for 50 heat bath sweeps before the first measurement.

The choice of the regions $P$ and $R$ is rather arbitrary; we chose to
divide the lattice by time planes at $T=1$ and $T=9$. We used two
different fuzzing levels for all operators.

A typical effective mass plot can be seen in
Figure~\ref{figure:effmass}, where we have plotted the effective mass
of the $L=0$ (S) state together with factorising fit.

\begin{figure}
  \centering
  \leavevmode
  \epsfxsize = 7.5cm
  \epsfbox{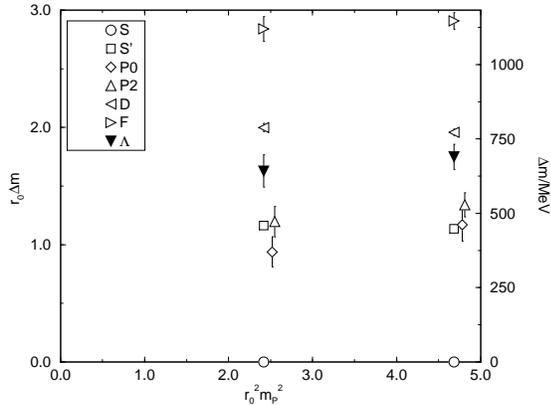}
  \vspace{-1cm}
  \caption{\protect The spectrum of B meson containing one light (with
mass proportional to pseudoscalar mass $m_P^2$) and one infinitely heavy
quark. Also included are our results for $\Lambda_b$.} 
  \label{figure:mB}
\end{figure}

Our full results can be seem from Figure~\ref{figure:mB},
where we have plotted the mass splittings between excited states and
ground state in units of $r_0$. Also included is the mass splitting of
$\Lambda_b$ and the ground state of $B$. Note that the ordering of P-waves we
observed before \cite{mp} has now changed. 

\begin{figure}
  \centering
  \leavevmode
  \epsfxsize = 7.5cm
  \epsfbox{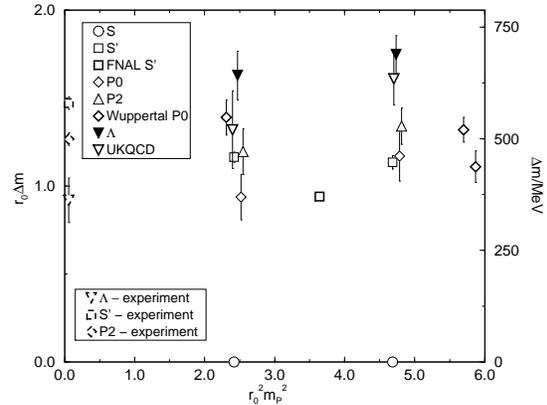}
  \vspace{-1cm}
  \caption{\protect Comparison with earlier results. FNAL group \cite{fnal}
has only estimated the excited S-wave with no error. UKQCD results for
$\Lambda_b$ are from \cite{ukqcd}, Wuppertal results for P0 from
\cite{Wuppertal}. The experimental values \cite{expt,expt2} are for real $B$
meson and $\Lambda_b$, so one would expect to see some difference from
leading order HQET.} 
  \label{figure:mBcomp}
\end{figure}

These results should be compared to results obtained using usual
inversion techniques. In Figure~\ref{figure:mBcomp} we have plotted
results several other groups \cite{fnal,ukqcd,Wuppertal} have
obtained using much 
more computing resources. Our results are clearly consistent with
theirs. However, we have much smaller errorbars and are able to obtain
reliably several excited states, which is not generally true for the
earlier work in the static limit. One should also note that our results
are compatible with results obtained from non-relativistic QCD \cite{khan}.

Because we have not yet estimated systematic errors arising from
finite lattice size and spacing, we chose not to extrapolate our results to
the chiral limit. 

\section{CONCLUSIONS}

We have shown that stochastic estimators for light quark propagators
are extremely useful in heavy quark effective theory, where the usual
method of calculating light quark propagators is extremely
wasteful. We have obtained reliable signals for the first time for excited
states in the static limit, and show that even states with $L=3$ (F-waves)
can be obtained. 

It is obvious that our method can be used successfully not only for
spectroscopy but also for matrix elements. The ability to obtain
reliable information on excited states is valuable when one is trying
to extract the heavy-light decay constant $f_B$.


\begin{thebibliography}{99}
\bibitem{divitiis} G. de Divitiis et al. Phys. Let., B382 (1996) 393.
\bibitem{hqet} See, for example, M. Neubert, Phys. Rept. 245 (1994)
259, and references therein. 
\bibitem{lm} UKQCD Collaboration, P. Lacock et al., Phys. Rev. D54
(1996) 6997. 
\bibitem{scale} UKQCD Collaboration, H.P. Shanahan et
al. Phys.Rev. D55 (1997) 1548  
\bibitem{mp} C. Michael and J. Peisa. hep-lat/9705013. 
\bibitem{fnal} A. Duncan et al. Phys. Rev. D51 (1995) 5101. 
\bibitem{ukqcd} UKQCD Collaboration, A. Ewing et al. Phys. Rev. D54
(1996) 3526. 
\bibitem{Wuppertal} C. Alexandrou et al. Nucl. Phys. B414 (1994) 815.
\bibitem{khan} For a review, see plenary talk by Arifa Ali Khan, these
proceedings. 
\bibitem{expt} C. Weiser. Proceedings of the 28th International
Conference on High Energy Physics, Warsaw 1996, p. 531. Ed. I. Adjuk
and A. Wroblewski, World Scientific 1996. 
\bibitem{expt2} R.M. Barnett et al., Phys. Rev. D54 1, (1996) 

\end{thebibliography}
\end{document}